\documentstyle[epsf, aas2pp4]{article}
\lefthead{Keohane, et al.}
\righthead{IC~443's Hard X-ray Emission}

\begin{document}

\title{A Possible Site of Cosmic Ray Acceleration  \\
           in the Supernova Remnant IC\,443}
\author{Jonathan W. Keohane\altaffilmark{1}, R. Petre\altaffilmark{2}, Eric V. Gotthelf\altaffilmark{3}}
\affil{The Laboratory for High Energy Astrophysics \\
       Code 662, NASA/GSFC, Greenbelt, MD \  20771}
\authoraddr{Code 662.0 \\ NASA/GSFC \\ Greenbelt, MD  20771}
\authoremail{jonathan@cassiopeia.gsfc.nasa.gov}
\author{M. Ozaki, K. Koyama}
\affil{Department of Physics, Faculty of Science \\ 
       Kyoto University, Sakyo-ku, Kyoto 606-01, Japan}

\altaffiltext{1}{Astronomy Department, The University of Minnesota; 
                 E-mail: jonathan@lheamail.gsfc.nasa.gov}
\altaffiltext{2}{E-mail: petre@lheavx.gsfc.nasa.gov}
\altaffiltext{3}{Universities Space Research Association;
                 E-mail: gotthelf@lheavx.gsfc.nasa.gov}
\bigskip  \bigskip
\centerline{February 5, 1997}
\bigskip
\centerline{Accepted by {\em The Astrophysical Journal}}

\begin{abstract}
We present evidence for shock acceleration of cosmic rays to high
energies ($\sim$10 TeV) in the supernova remnant IC\,443.  X-ray
imaging spectroscopy with {\it ASCA} reveals two regions of
particularly hard emission:  an unresolved source embedded in an
extended emission region, and a ridge of emission coincident with the
southeastern rim.  Both features are located on part of the radio shell
where the shock wave is interacting with molecular gas, and together
they account for a majority of the emission at 7 keV\@.  Though we
would not have noticed it {\it a priori}, the unresolved feature is
coincident with one resolved by the {\it ROSAT} HRI.  Because this
feature overlaps a unique region of flat radio spectral index ($\alpha
< 0.24$), has about equal light-crossing and synchrotron loss times,
and a power law spectrum with a spectral index of $\alpha$=1.3$\pm$0.2,
we conclude that the hard X-ray feature is synchrotron radiation from a
site of enhanced particle acceleration.  Evidence against a plerion
includes a lack of observed periodicity (the pulsed fraction upper
limit is 33\%),  the spectral similarity with the more extended hard
region, the location of the source outside the 95\% error circle of the
nearby {\it EGRET} source, the fact that it is nestled in a bend in the
molecular cloud ring with which IC\,443 is interacting, and the
requirement of an extremely high transverse velocity ($\ge$5,000
km/s).  We conclude that the anomalous feature is most likely tracing
enhanced particle acceleration by shocks that are formed as the
supernova blast wave impacts the ring of molecular clouds.
\end{abstract}


\section{Introduction}

  It is generally accepted from energy budget considerations that
supernova remnants (SNRs) are the primary source of galactic cosmic
rays (CRs), which are inferred to span the energy range from about
$10^8$ to $10^{14}$ eV (Blandford \& Ostriker 1978; Axford 1994;
Biermann 1995, and references therein)\@.  Though radio observations of
synchrotron radiation from supernova remnants supply bountiful evidence
that they are the source of CRs at GeV energies, evidence regarding
higher energy cosmic rays is largely circumstantial.  A search for
ultra-high energy $\gamma$-rays ($\sim$10$^{14}$ eV) coincident with
SNRs detected nothing significant (Allen, et al.\ 1995)\@.  The
best evidence thus far for shock acceleration of $\sim$100 TeV CRs,
comes from X-ray observations of SN\,1006 (Koyama et al.\ 1995)\@.
SN\,1006 appears to be unique in that the non-thermal X-ray flux from
its shell dominates the thermal, but current limits on the non-thermal
flux from the shells of other young SNR do not exclude the presence of
a synchrotron component, and hence ongoing CR acceleration to TeV
energies.

In this paper we report the discovery of a site in another SNR in which
high energy cosmic rays are possibly being accelerated, but by a
different mechanism from that in SN\,1006.    Using {\it ASCA} GIS
data, we have found a concentration of hard X-ray emission along the
southern rim of the middle-aged remnant IC\,443, whose spectrum is
consistent with a power law, and whose morphology suggests enhanced
shock acceleration resulting from the SNR shock encountering dense
clouds in the ISM.  The observations are consistent with the prediction
of such an effect by Jones \& Kang (1993, hereafter JK93)\@.

IC\,443 (G189.1+3.0) is a nearby ($\approx 1.5$ kpc) supernova remnant
(SNR) of intermediate age.  Many infrared and radio line observations
have demonstrated that IC\,443 is interacting with a ring of molecular
clouds (e.g., DeNoyer 1979; Burton {\it et al.}\ 1988, hereafter BGBW;
Burton {\it et al.}\ 1990; Dickman {\it et al.}\ 1992,
hereafter DSZH; van Dishoeck {\it et al.}\ 1993)\@. The region in which
the most complex interactions between the molecular cloud and the SNR shock
is occurring also has an unusually flat radio spectral index for a
shell-like SNR ($\alpha<$0.24 -- Green 1986, hereafter G86)\@.

In the X-ray band, IC\,443 has been the subject of a number of
comprehensive studies, most notably those of Petre {\it et al.}\ (1988,
hereafter PSSW) and Asaoka \& Aschenbach (1994, hereafter AA94)\@.  It
has an irregular soft X-ray morphology, influenced strongly by its
interactions with an H I cloud to the northeast and the foreground
molecular cloud and shadowing by the foreground SNR G189.6+3.3.  Using
{\it Ginga}, Wang et al.\ (1992, hereafter WAHK) resolved spectrally a
hard component which they were able to characterize by either a thermal
model with kT$\sim$14 keV or a power law with photon index
$\Gamma\sim$2.2.  Scanning observations by {\it HEAO1} A-2 showed that
the hard emission is more centrally located than the bright soft
emission. Neither the A-2 nor the {\it Ginga} observations provided
information regarding the extent of the hard emission.

IC\,443 is also coincident with a high energy $\gamma$-ray source,
which has led to speculation that the $\gamma$-rays are produced by the
interaction of cosmic rays, accelerated by IC\,443's shocks,  with
nearby molecular cloud material (Sturner \& Dermer 1995; Esposito {\it
et al.}\ 1996, hereafter EHKS)\@.  Models of broad-band non-thermal spectra
have recently been produced by Sturner et al.\ (1996); in these
synchrotron radiation dominates from the radio to the soft X-ray, while
electron bremsstrahlung and $\pi ^{0}$ decay dominate the $\gamma$-ray
emission for a supernova remnant like IC\,443.

Our paper is organized as follows.  We first present the {\it ASCA} GIS
images showing the location of the hard X-ray emission region
(\S\ref{ASCA_maps.sec})\@.  We compare the {\it ASCA} image with that
from the {\it ROSAT} HRI\@.  We then discuss our analysis of the GIS
spectrum and the GIS and HRI light curves from that region
(\S\ref{ASCA_spectra.sec})\@.  Finally, we discuss possible emission
mechanisms and the implications for the acceleration of cosmic rays by
supernova remnants (\S\ref{discussion.sec})\@.

\section{Observations and Analysis \label{analsis.sec}}

\begin{figure}[htb]
\epsscale{1.00}
\plottwo{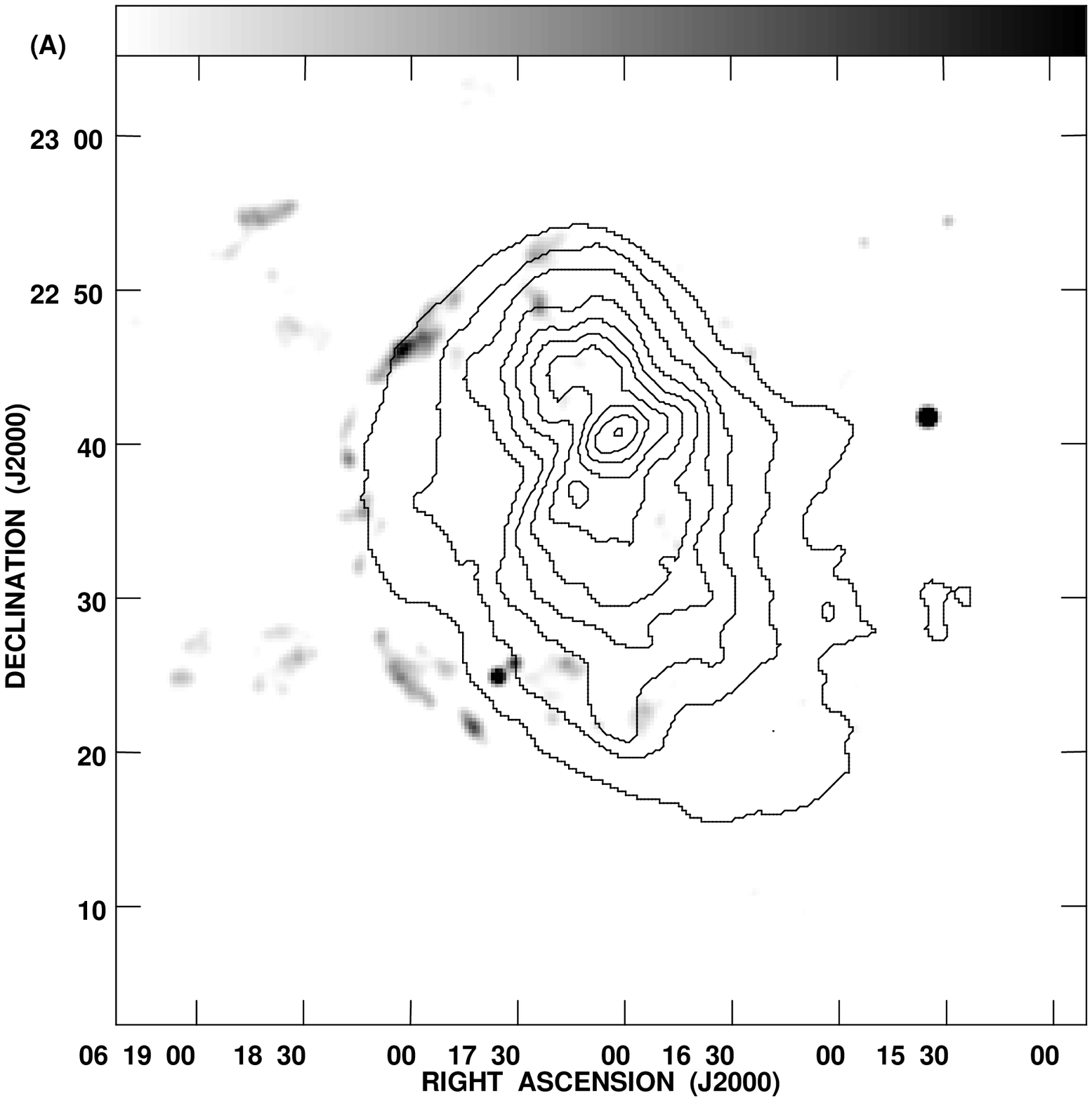}{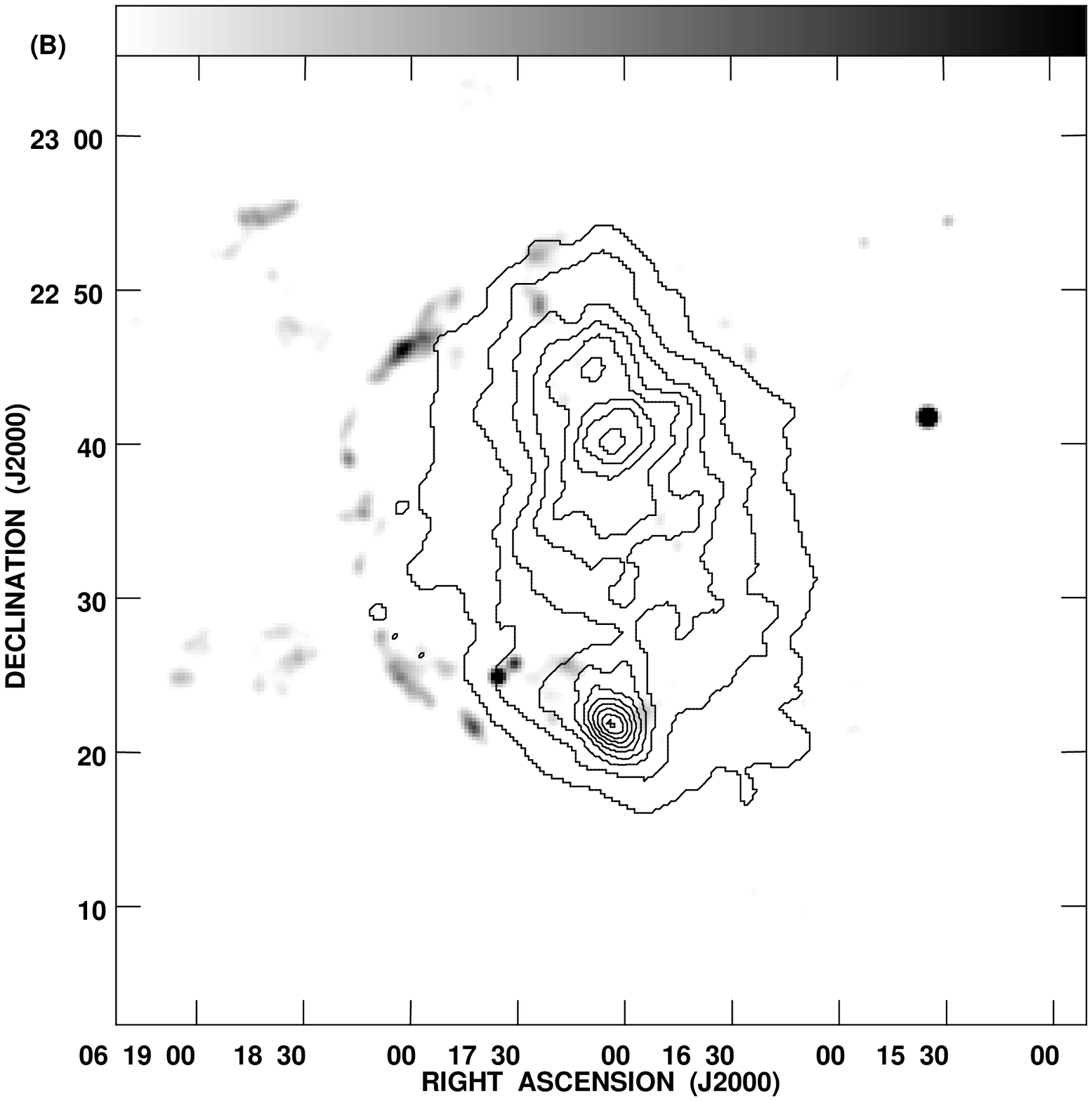} \\
\plottwo{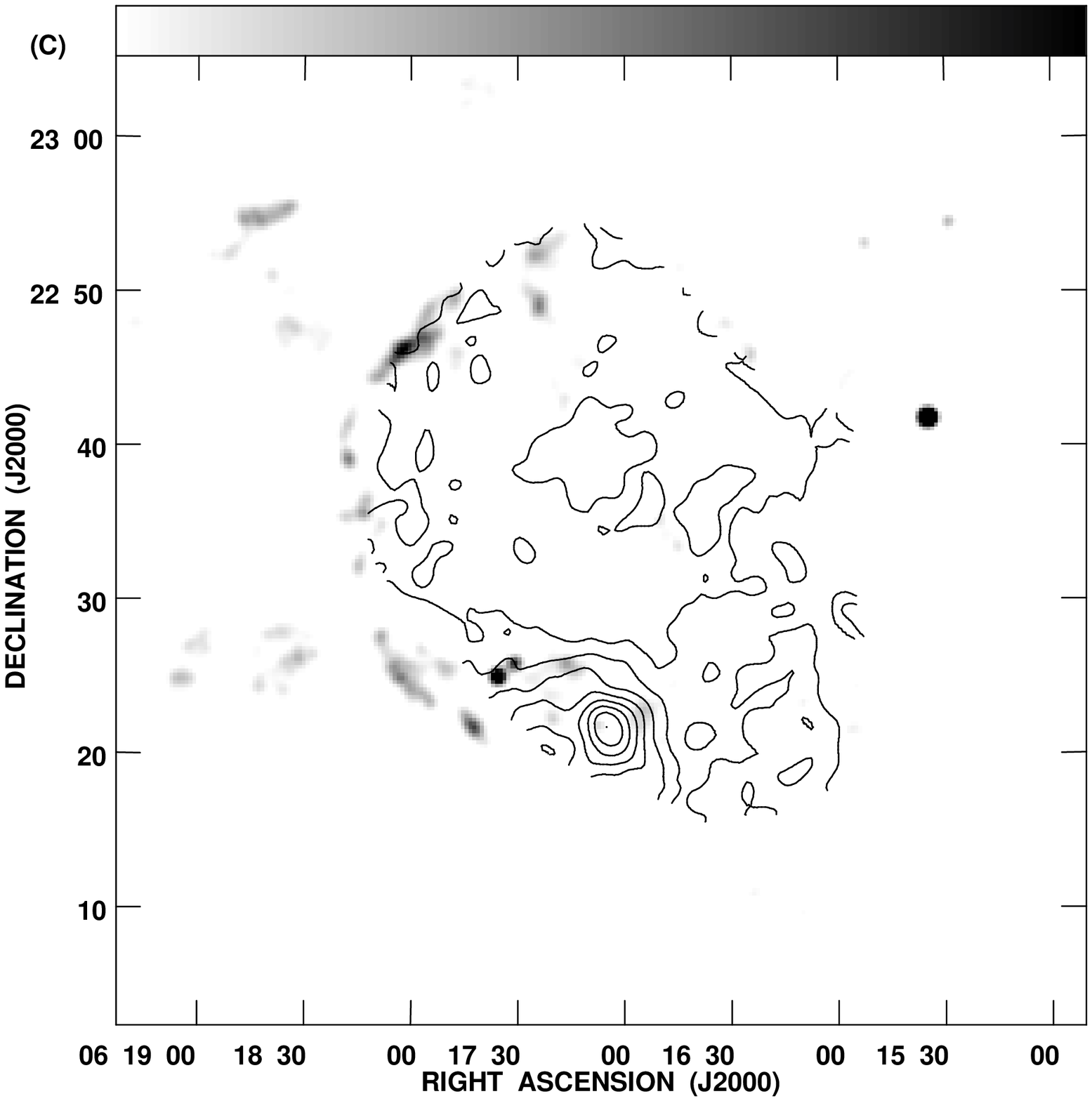}{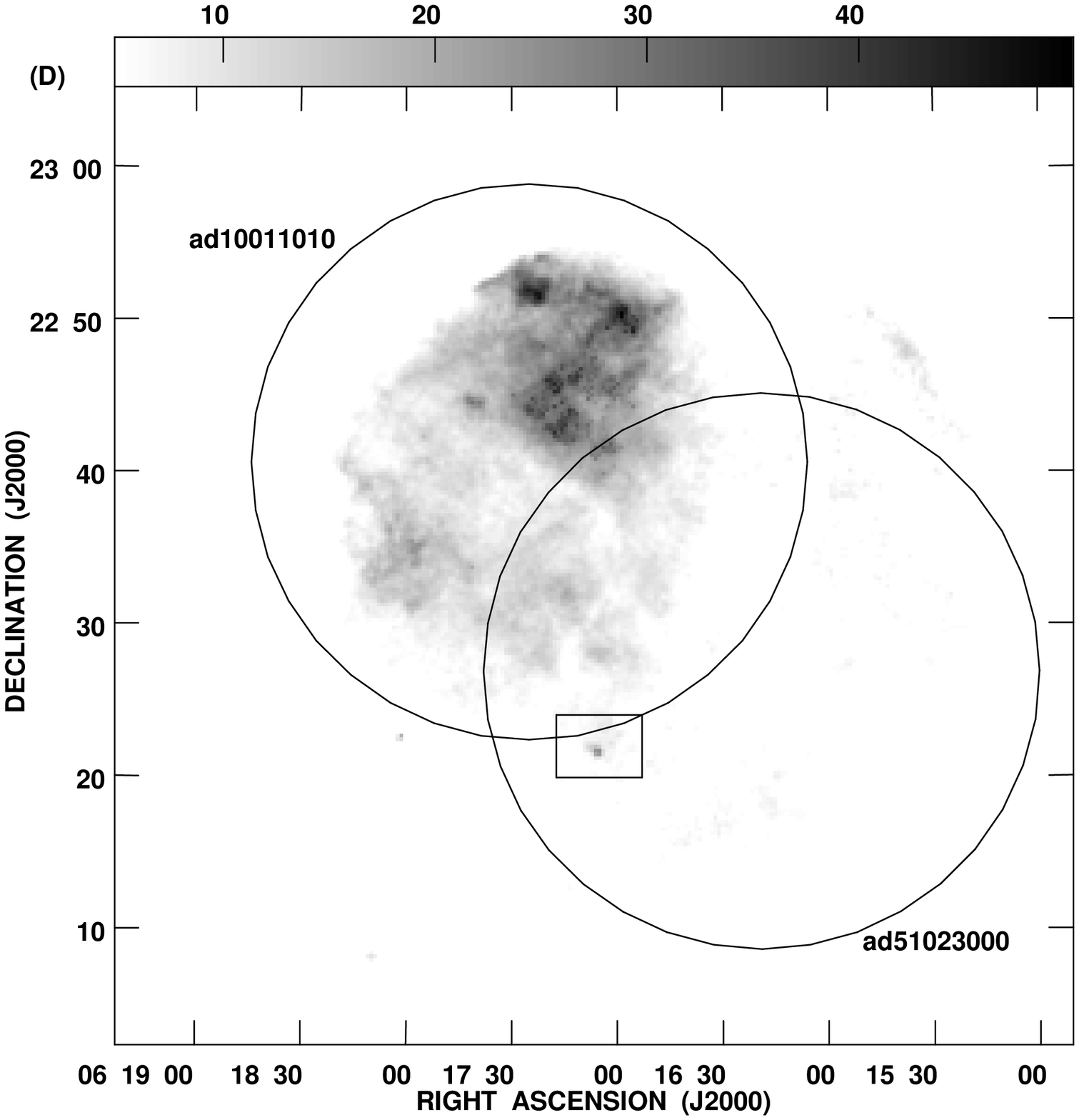} \\
\epsscale{1.00}
\caption{\label{asca_images}Soft (A) and hard (B) GIS contour images of
IC\,443.  The energy bands are 1.1--2.1 and 2.1--12.0 keV
respectively.  The ratio of the two bands (C) is also shown (blanked in
regions of low soft X-ray surface brightness).
 The contour levels in each figure are multiples of 10\% of the peak.
For reference, we have shown in grey-scale (A, B, C), a 1.4 GHz radio
map from the VLA sky survey (10--100 mJy/beam)\@.  Image D shows our
{\it ROSAT} HRI mosaic of IC\,443; the GIS field of view for each
observation and the region shown in Figure \ref{hri_image} are
overlaid.}
\end{figure}

\begin{table} [hbtp]
\caption{ASCA observations of the SNR IC\,443  \label{obs_tab}}
\begin{tabular}{||l|c|c||}  \hline  \hline
Sequence \# & 10011010 & 51023000 \\
Mission Phase    & PV phase & AO-1 \\
RA (J2000)  &  $\rm 06^h 17^m 25^s$ &  $\rm 06^h 16^m 19^s$  \\
Declination & +22\arcdeg 40\arcmin 37\arcsec & +22\arcdeg 26\arcmin 53\arcsec \\GIS Exposure&  20.2 ks & 34.7 ks \\  
Date        & April 14, 1993 & March 9, 1994 \\ \hline \hline 
\end{tabular}
\end{table}

\begin{figure}[hbt]
\epsscale{1.00}
\plotone{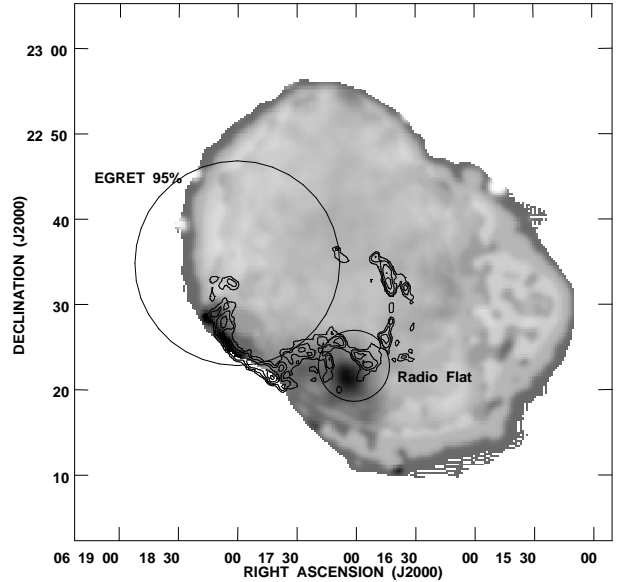} \\
\epsscale{1.00}
\caption{\label{figure_discussion.fig} The {\it ASCA} GIS hardness ratio map
($F_{\mbox{\tiny 2.1--12 \rm keV}}/F_{\mbox{\tiny 1.1--2.1 \rm keV}}$) is
shown in grey-scale.  The overlaid drawings are:  the EGRET $\gamma$-ray
detection circle (EHKS), the region of flat radio spectral index (G86)
and a contour plot of the v=1-0S(1) $\rm H_2$ emission line intensity
(BGBW)\@.}
\end{figure}

\subsection{The {\it ASCA} GIS X-ray Maps \label{ASCA_maps.sec}}

{\it ASCA} performed observations of two adjacent regions of IC\,443,
one during the Performance Verification (PV) phase of of the mission,
the other during the first cycle of guest observations (AO-1)\@.  The
relevant information about these observations is contained in Table
\ref{obs_tab}\@.  We extracted these data sets from the {\it ASCA}
public archive.  In figures \ref{asca_images} A and B, we show
exposure-corrected GIS mosaic images for the hard (2.1--12 keV) and
soft bands (1.1--2.1 keV)\@.  These appear highly correlated, except
for a bright feature in the hard band map, centered at $\rm RA = 06^h
17^m 05^s$, $\rm Dec = +22\arcdeg 21\arcmin 30\arcsec$ (J2000)\@.  The
anomalous nature of this feature is shown in dramatic fashion in figure
\ref{asca_images}C, a spectral hardness ratio map, constructed by
dividing the hard band map by the soft.  There it can be seen that the
other features, associated with the diffuse thermal emission in IC 443,
all have similar spectral hardness, but the hard X-ray feature (HXF)
stands out.  The {\em ASCA} GIS brightness distribution within the hard
feature is consistent with that expected from a source smaller than
$\sim 1 \arcmin$\@.  The hard feature is $\sim$12$\arcmin$ off axis so
it was not in the SIS field of view.

\begin{figure}[hbt]
\epsscale{1.0}
\plotone{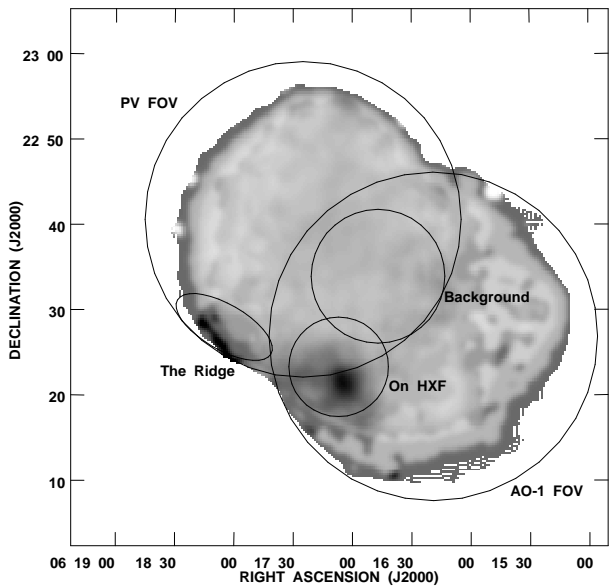} \caption{\label{spectra_region} The
regions where {\it ASCA} GIS data were selected for our spectral
analysis and the approximate {\it ASCA} GIS fields of view, overlaid on
the same hardness ratio map as shown in
Fig.~\ref{figure_discussion.fig}.  We purposely chose a larger than
needed region about the HXF and the ridge in order to spectrally
distinguish between the thermal and non-thermal spectra.}
\end{figure}

In addition, there is a ridge of spectral hardness located northeast of
the HXF along the radio-bright rim of the remnant.  This second feature
is not shown in Fig.\ \ref{asca_images}C because it was blanked due to
its small soft X-ray count rate.  It is shown, however, in
Figs.\ \ref{figure_discussion.fig} and \ref{spectra_region}\@.  As will
be discussed in more detail in \S\ref{ASCA_spectra.sec}, the hard flux
from the observed portion of the ridge is about half the hard flux from
the HXF and the two regions can account for a majority of the total
hard ASCA flux\@.

\begin{figure}[hbt]
\epsscale{1.0}
\plotone{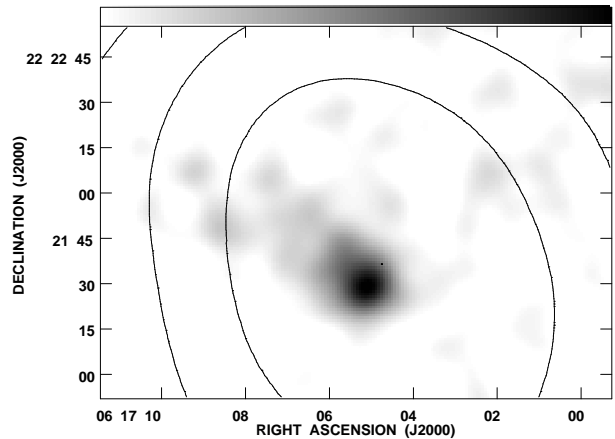} 
\epsscale{1.0}
\caption{\label{hri_image}  The 10\arcsec\ FWHM Gaussian smoothed {\it
ROSAT} HRI (grey-scale) and the 60\arcsec\ resolution hardness {\it
ASCA} hardness ratio map shown with the same contour levels as figure
\ref{asca_images}C\@.  The grey-scale range is 10-45 counts per
10\arcsec\ diameter circular ``beam.'' }
\end{figure}

\subsection{{\it ROSAT} HRI Image \label{rosat_hri_image.sec}}

The region containing the HXF was observed for 30 ks using the {\it
ROSAT} HRI as part of a program to create a complete high resolution
X-ray  image of IC\,443.  In Fig.\  \ref{asca_images}D, we show the
complete HRI mosaic, with the GIS fields of view overlaid.  A surface
brightness enhancement appears approximately at the location of the
HXF\@.  The HRI map of the region containing the HXF is shown in Fig.\
\ref{hri_image}, with the GIS hardness ratio contours overlaid.  The
feature's central core has an angular extent of about
$10\arcsec$$\times$$5\arcsec$, with extended lower level structure of
about 1\arcmin\ in extent.  The elongation of the central core is
perpendicular to the orientation of the low level emission and does not
resemble the HRI point spread function, so we believe it to be
resolved.

\subsection{Spatial Correlations with Other Bands \label{spatcor.sec}}

In Fig.\  \ref{figure_discussion.fig} we overlay on the GIS spectral
hardness ratio map possibly related spatial information from other
bands:  the EGRET 95 percent confidence error circle (EKHS), contours
of H$_2$ emission (BGBW) which locate the sites of the most intense SNR
shock/cloud interaction, and the region in which the
$\lambda\lambda$20--200 cm spectral index is appreciably flatter than
elsewhere in IC\,443 (G86)\@.  The HXF is just outside a
concave arc of H$_2$ emission, but correlates well with the region of
flat radio index.  Interestingly, it is well outside the EGRET error
circle.  In the Digitized Sky Survey (Lasker et al.\ 1990), there are
some faint filaments in this region, but no obvious unresolved sources
near the hard feature.  A search of the most recent on-line pulsar
catalog (Taylor et al.\ 1993) reveals none near the feature.

In addition to the main hard feature, the ridge of hard emission
overlaps the H$_2$ emission region, but does not appear coincident with
any region of radio spectral index flattening.  However, it does lie
along the bright radio rim.

\begin{table*} [t]
\renewcommand{\thefootnote}{\alph{footnote}}
\setcounter{footnote}{0}
\caption{Spectral Fits to Background Field and Hard X-ray Features  \label{good_fit.tab}}
\begin{tabular}{||rlcccccc||}  \hline  \hline 
&Model & Data Sets & $\chi^{2}$ & $\chi^{2}_{\nu}$ &$T_{\rm soft}$ & $N_{\rm H}$ & Hard \\ 
& & & & & keV &  $10^{22}$cm$^{-2}$ & Component \\ \hline
1&Thermal\footnotemark[1]& Background & ~781  &  0.71 & 0.89 $\pm$ 0.03 & 0.12 $\pm$ 0.03 & {\sc n/a} \\
2&Thermal\footnotemark[1] & On HXF & ~793  &  0.73 & 1.9 $\pm$ 0.1 &  0 & {\sc n/a} \\
3&Thermal\footnotemark[2]& On HXF & 1407  &  1.28 & 0.89 & 0.31 $\pm$ 0.03 & {\sc n/a} \\
4&Power-law\footnotemark[3] + Thermal\footnotemark[2] & On HXF & ~769 & 0.70 & 0.89 & 0.18 $\pm$ 0.03 & $\Gamma = 2.3 \pm 0.2$ \\
5&Power-law + Thermal\footnotemark[4] & On HXF & ~766 & 0.70 & 1.1$^{+0.3}_{-0.1}$ & 0.12 & $\Gamma = 2.4^{+0.2}_{-0.4}$ \\
6&Hot Bremss + Thermal\footnotemark[2] & On HXF & ~763 & 0.70 & 0.89 & 0.13 $\pm$ 0.03 & kT=$3.9_{-0.6}^{+1.7}$ keV \\
7&Thermal\footnotemark[4]& The Ridge & ~778  & 0.40 &  0.74 $\pm$ 0.08 & 0.12 & {\sc n/a} \\
8&Power-law + Thermal\footnotemark[2] & The Ridge & ~755 & 0.39 & 0.89 & 0$^{<0.03}$ & $\Gamma = 0.1 \pm 1.3$ \\
9&Power-law + Thermal\footnotemark[4] & The Ridge & ~748 & 0.38 & 0.54$^{+0.07}_{-0.16}$  & 0.12 & $\Gamma = 1.6^{+1.4}_{-0.7}$ \\
 \hline \hline 
\end{tabular}  \\
Fits are to the energy range 0.5--12 keV  \\
\footnotemark[1] Letting the scaling (GIS2 \& GIS3), column density,
temperature and line strengths vary \\
\footnotemark[2] We froze the line strengths and temperature to the off
source (model \#1) values, but fit the column density and scaling. \\
\footnotemark[3] The best-fit 5 keV flux density is $\left( 7 \pm 3 \right) \times 10^{-13} \frac{\rm erg}{\rm s \, cm^{2} \, keV}$\@.\\
\footnotemark[4] We froze the column density and line strengths to the off
source (model \#1) values, but fit the temperature and scaling. \\
\end{table*}

\subsection{The {\it ASCA} Spectrum \label{ASCA_spectra.sec} }

The HXF and the ridge are embedded within the
diffuse emission of IC\,443 --- at least in projection.  Thus in order
to determine the spectral properties of these features, one must first
adequately characterize the ``background'' thermal emission.  We
therefore extracted a spectrum from the adjacent larger region shown
in figure~\ref{spectra_region}.  We assume, based on the smoothness of
the spectral hardness ratio map, that the spectrum of the diffuse
emission does not vary appreciably in the neighborhood of the HXF;
based on the previous X-ray observations the spectral variations in the
diffuse emission would be most significant below 1 keV, outside the
effective GIS band.  To this ``background'' spectrum we fit an {\it ad
hoc} thermal emission model, comprised of a thermal bremsstrahlung
continuum and narrow Gaussians to represent the most prominent emission
lines (the He$\alpha$ transitions of Ne, Mg, Si, S and Ar)\@.   Such 
{\it ad hoc} models have been used previously to model
{\it ASCA} spectra from other remnants (e.g. Miyata {\it et al.}\ 1994;
Holt {\it et al.} 1994)\@.  The best fit model yields kT$\sim$0.9 keV,
comparable with published temperatures for IC\,443 (PSSW, WAHK)\@.
However, it is important not to over-interpret our thermal results,
because we have made no attempt to distinguish the He$\alpha$
transitions from weaker nearby spectral lines (including each element's
corresponding H-like Ly$\alpha$ transition) or to develop a physically
self-consistent thermal model\@.

\begin{figure}[bt]
\epsscale{1.0}
\plotone{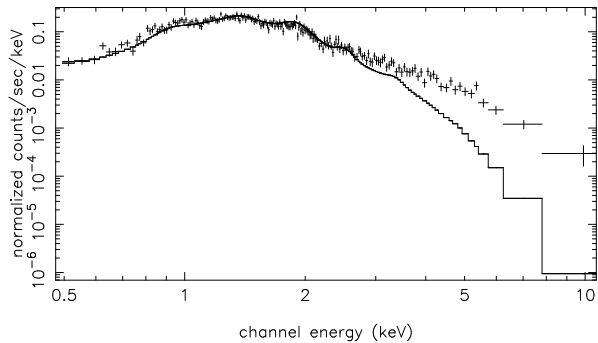}
\caption{\label{spectra} The {\it ASCA} GIS2 spectrum of the hard feature in
IC\,443 with the best-fit {\em ad hoc} thermal model.  By adding a hard
power law component to this model we reduce the combined GIS2 and GIS3 
$\chi^2$ by $\Delta \chi^2 = 640$\@. }
\end{figure}

We next applied this same model to the GIS spectrum of the region
containing the HXF (also shown in figure~\ref{spectra_region}), allowing
only the normalization to vary.  As shown in table \ref{good_fit.tab},
this yields an unacceptable fit. As indicated in Fig.\
\ref{spectra}, the ``background'' model provides a good fit to
the data below $\sim$2 keV, but there is a substantial flux excess at
higher energies.  An acceptable fit to the ``on HXF'' spectrum is
obtained by adding a second continuum component.  The current data do
not allow us to distinguish between a power law with
$\Gamma$=2.3$\pm$0.2 and thermal bremsstrahlung with 
kT=$3.9_{-0.6}^{+1.7}$~keV.

We also performed this same analysis using a smaller on-source region,
tightly drawn around the HXF\@.   However, this was less effective:
because of the lack of thermal background counts, it is harder to
spectrally distinguish the background from the non-thermal source
emission.  Nevertheless we measure a photon index of
$\Gamma$=$2.2_{-0.4}^{+0.1}$, using the technique described above.

We applied the same spectral analysis procedure as described above for
the HXF to the hard ridge\@.  Since the ridge is farther from the
``background'' region than the HXF (see Fig.~\ref{spectra_region}) and
we fit the thermal model using the AO-1 data, we were uncertain if it
was appropriate to use the same background thermal model for both the
HXF and the ridge.  To test this, we performed our analysis using both
the AO-1 ``background'' region shown in Fig.~\ref{spectra_region} and
one using PV phase data closer to the ridge.  We do not detect any
significant difference in the spectral index or flux of the hard ridge
region as a function of the chosen background region.

Because of its larger extent and lower surface brightness, the ridge
spectrum has lower signal to noise than the HXF\@.  As with the HXF,
fitting the spectrum with the background model yielded a clear excess
above 2 keV\@.  It was not possible, however, to constrain the spectral
properties of the ridge's hard component.  While we have no evidence
suggesting a different spectral form, we cannot eliminate the
possibility that the hard emission from the ridge has a different
origin.

The GIS spectral parameters for this region are similar to those
obtained using the non-imaging (1\arcdeg field of view) {\it Ginga}
LAC\@.  WAHK reported the presence of a hard component that was fit by
a thermal model with kT=11$^{+7.5}_{-2.0}$ keV (from their Fig.\  3) or
a power-law with photon index $\Gamma=2.2\pm0.13$\@.  Assuming a uniform
surface brightness inside the shell and an absorbing column density of
$N_{\rm H} = 10^{21.9} \, {\rm cm^{-2}}$, WAHK found the total 2--20
keV emitted flux from IC\,443 to be $9 \times 10^{-11} \, {\rm erg \,
cm^{-2} \, s^{-1} }$\@.  As the total flux is dominated by the $\sim$ 1
keV thermal component, a comparison between the broad band {\it Ginga}
flux and the GIS flux from the hard features is inappropriate.  A more
reasonable quantity to compare is the flux density at 7 keV, where the
relative contribution of the $\sim$ 1 keV component is negligible.
Using Figure 2 in WAHK, and assuming that the {\it Ginga} LAC has an
effective area of 3000 cm$^2$ and near unity quantum efficiency at 7
keV, we estimate the {\it Ginga} flux density at 7 keV to be 20
$\times$ 10$^{-5}$ photons~s$^{-1}$~keV$^{-1}$~cm$^{-2}$\@.  From the
GIS data, we find the HXF flux density at 7 keV to be approximately 4
$\times$ 10$^{-5}$ photons~s$^{-1}$~keV$^{-1}$~cm$^{-2}$, and the flux
density of the portion of the hard ridge within the field of view to be
about 2 $\times$ 10$^{-5}$ photons~s$^{-1}$~keV$^{-1}$~cm$^{-2}$\@.  Thus
the detected flux in the hard regions seems to account for only 30\% of the
hard flux detected by {\it Ginga}, despite their prominence in the GIS
images.  Moreover, integration over the rest of the surface of the
remnant contained in the GIS fields (about 90 percent of the total)
yields a flux density of no more than 4 $\times$ 10$^{-5}$
photons~s$^{-1}$~keV$^{-1}$~cm$^{-2}$, leaving a factor of two
discrepancy between the GIS and {\it Ginga} 7 keV flux densities.

Two possible resolutions of this discrepancy are that there is
significant hard flux from the small fraction of IC 443 unobserved by
the GIS, or there is a calibration discrepancy between the GIS and {\it
Ginga}.  To check this latter possibility, we compare the broad band
flux from all of IC 443 in the common 2-10 keV band.  The 2-20 keV {\it
Ginga} flux is 9 $\times$ 10$^{-11}$ ergs~cm$^{-2}$~s$^{-1}$ (WAHK).
The 2-10 keV flux from IC 443 as measured by {\it HEAO1} A-2 is 7$\pm$1
$\times$ 10$^{-11}$ ergs~cm$^{-2}$~s$^{-1}$ (PSSW), consistent with the
{\it Ginga} measurement.  Integrating the 2-10 keV flux within the
field of view of the two GIS pointings, we find 5$\pm$1 $\times$
10$^{-11}$ ergs~cm$^{-2}$~s$^{-1}$; this represents 90 percent or more
of the total flux from the remnant.  The consistency of these numbers
indicates that contribution of cross calibration uncertainties to the
discrepancy is probably small, but could be as large as a factor of 2.
Thus while it is possible that the GIS has observed all the regions in
IC 443 producing hard flux, we cannot rule out the possibility that the
unobserved portion of the IC 443 rim to the south and east of the ridge
contribute significantly to the hard flux.

A key conclusion about the spatial distribution of the hard X-ray
emission in IC 443, independent of a resolution of this discrepancy, is
that the hard emission is highly localized.  The flux density from the
bulk of the remnant surface area is at most 40 percent of the total
hard flux.  This provides a strict upper limit on the luminosity of a
hard thermal component, distributed throughout the remnant.  WAHK
inferred that IC\,443 is a very young (1200 y) remnant based on their
interpretation of the hard component as hot gas arising from a high
velocity shock, and IC\,443's coincidence with the guest star of 837
A.D\@.  Despite the restrictive limit we have placed on the flux of a
hot thermal component, we nevertheless cannot rule out their conclusion
about the age of IC\,443\@.  Moreover, high shock velocities may also
be required to accelerate electrons to the relativistic energies
required for the emission of X-ray synchrotron radiation, and WAHK's
historical evidence is naturally unaffected by our observations.

\subsection{Pulsation Search \label{periodicity.sec}}

In order to search for periodicity from the unresolved HXF, we
performed coherent fast Fourier transforms (FFTs) on the {\it ROSAT}
HRI and {\it ASCA} events recorded from the region used for spectral
analysis (after applying the standard barycenter corrections).  All our
results are consistent with a non-periodic signal.

The temporal resolution of the GIS is 0.0625 s, so we cannot use the
GIS data to search for a frequency faster than 8~Hz.  On the other
hand, the {\it ROSAT} HRI data have millisecond resolution.  We
therefore  performed summed FFTs on the combined data set ({\it ROSAT}
+ {\it ASCA}) in the low frequency range 0.05--8 Hz, and only the {\it
ROSAT} data in the high frequency range 10--1000 Hz.  We find 99\%
confidence pulsed fraction upper limits of $\sim$15\% and $\sim$33\% in
the low and high frequency ranges respectively.

\section{Discussion}
\subsection{Summary of Results \label{obs_facts.sec}}

We have discovered an isolated hard X-ray emitting feature and a ridge
of hard emission in the southeast of the SNR IC\,443 using the {\it
ASCA} GIS\@.  The HXF's X-ray spectrum can be characterized by either a
power law of energy spectral index $\alpha$=1.3$\pm$0.2 or thermal
bremsstrahlung with kT=$3.9_{-0.6}^{+1.7}$ keV; the ridge spectrum
appears similar. The features can account for most of the hard X-ray
flux from IC443 (\S\ref{ASCA_spectra.sec})\@.  The core of the HXF is
marginally resolved with the ROSAT HRI, but the low level surrounding
emission extends about 10\arcmin\ along the radio-bright shell.
(\S\ref{rosat_hri_image.sec}, fig.  \ref{hri_image})\@.  It is
spatially coincident with the $\lambda\lambda$20--200~cm flat spectral
($\alpha \approx 0.2$) region (G86; fig. \ref{figure_discussion.fig}),
and in the general region of, but not uniquely coincident with, regions
of cloud/shock interactions (DeNoyer, 1979; fig.
\ref{figure_discussion.fig})\@.  It is located outside the 95 percent
confidence error circle of the nearby EGRET source (EHKS; fig.
\ref{figure_discussion.fig})\@.  No strong periodicity is observed from
the feature, with a pulse fraction upper limit of 33\%
(\S\ref{periodicity.sec}), and there is no known pulsar near it.

\subsection{Origin of the Emission \label{discussion.sec}}

We focus here on the origin of the emission from the HXF\@.  Because of
the uncertainty of the spectral parameters we know substantially less
about the ridge.  While some of our discussion below applies to all the
emission (e.g. our conclusion that it is non-thermal), a more
definitive understanding of the nature of the ridge emission (which
could be different from that of the HXF) awaits a better quality X-ray
spectrum.

The data suggest that the most
likely mechanism for producing hard X-rays is synchrotron emission.
Before addressing whether this emission arises from a plerion or shock
accelerated electrons, we first discuss our rationale for ruling out
alternative mechanisms, including bremsstrahlung from a thermal and
a non-thermal population of electrons, and inverse-Compton scattering.

\paragraph{Thermal Bremsstrahlung}  A thermal bremsstrahlung model
provides an adequate fit to the {\it ASCA} and the {\it Ginga} spectra
separately.  The inferred temperatures are discrepant, however
($\approx15$ keV for {\it Ginga} versus $\approx4$ keV for {\it
ASCA})\@.  In contrast, a power law model yields the same photon
index.  If the 2--20 keV spectrum is better characterized by a
power law, then the temperature obtained from a thermal model fit from
each instrument would yield a temperature characteristic of the
instrumental bandpass, which is exactly the case here.  Thus, a thermal
bremsstrahlung model is inconsistent with the {\em ASCA} and {\em
GINGA} observations taken together.  

\paragraph{Inverse Compton}  One possibility discussed by WAHK is that
X-rays are produced by inverse-Compton scattering of infrared photons
off electrons.  Gaisser et al.\ (1996) found that an inverse-Compton
component, scattering off the microwave background, would have a photon
index of $\Gamma$=$1.5$\@.  On the other hand, if the scattering
photons were locally produced, one would expect the hard X-ray emission
to arise from IR-bright regions.  The localization of the hard X-ray
emission a restricted region, the brightest part of which has no IR
counterpart argues against an inverse-Compton origin.

\paragraph{Accelerated Bremsstrahlung}  
It has been suggested that electrons accelerated in the shock to MeV
energies may be responsible for the generation of hard X-rays via
bremsstrahlung in supernova remnants (Sturner {\it et al.}\ 1995;
Asvarov {\it et al.}\ 1990)\@.  The fact that the radio spectral index
around the hard X-ray feature is flatter than elsewhere in IC\,443 is
strong evidence against this mechanism.  If we assume that the same
acceleration mechanism produces both the MeV and the GeV electrons
(responsible for the radio emission), then there will be fewer MeV
electrons in the hard region than elsewhere in the remnant.  Thus, if
this mechanism were operating efficiently, this region would be {\em
dimmer} in hard X-rays than everywhere else in the remnant.

\paragraph{Synchrotron Radiation}

The most straightforward mechanism for producing a power law X-ray
spectrum is synchrotron radiation.  None of the observations contradict
this interpretation; all are consistent with it.  In particular, the
fact that the radio spectrum of this region is flatter than elsewhere
argues that at higher energies (including the X-ray band), the
synchrotron flux will be enhanced over elsewhere in the SNR.  In fact,
as we show below, the physical size of the HXF corresponds well with
the synchrotron loss time of X-ray producing electrons.

WAHK dismissed synchrotron emission as a likely model, because an
acceleration mechanism efficient enough to accelerate electrons to
$\sim$10 TeV seemed unlikely and they calculated synchrotron loss times
to be less than the age of the SNR.  Their dismissal was based on the
assumption that the hard X-ray emission is spatially uniform; we now
know from this {\it ASCA} observation that the hard emission arises
primarily from localized regions.  

X-ray synchrotron emission in a SNR can be produced by high energy
electrons accelerated and interacting with a magnetic field in one of
two locations:  the magnetosphere of a pulsar (giving rise to a
plerion) or the forward shock of a SNR.  The observations support both
interpretations to some extent, but more issues arise from the presence
of a plerion than from an isolated region of intense shock
acceleration.

The compact size of the hard X-ray feature is the primary evidence in
favor of a plerionic interpretation.  The lack of X-ray pulsations is
not necessarily a problem:  the upper limit on the pulsed fraction is
higher than other pulsars like Vela (Pravdo  et al.\ 1976)\@.  Nor is
the existence of extended emission without an obvious embedded point
source unprecedented:  the SNR 3C 58 has a small extent, a power law
spectrum and shows no pulsations, but is generally regarded as a plerion
(Helfand et al.\ 1995)\@.    

Difficulties with the interpretation arise when trying to associate a
plerion with IC\,443.  Wilson (1986) has shown that the median X-ray:radio $\left(\frac{\mbox{0.5--3.5 keV}}{\mbox{$10^7$--$10^{11}$ Hz}} \right)$
flux ratio for plerions is about 1, with a range from
about 0.1--500\@.  The flux ratio of the HXF is about 900,
which would give it the highest known X-ray:radio flux ratio of any plerion.

Furthermore, if the feature is a plerion associated with the IC\,443
shell it requires an extremely high projected velocity of $5,000 \left(
\frac{d}{1.5 \, {\rm kpc}} \right) \left( \frac{\theta}{10\arcmin}
\right) \left( \frac{t} {1000 \, {\rm yr}} \right)$ ${\rm  km \,
s^{-1}}$, for distance $d$, angular distance $\theta$ from the
explosion center, and age $t$\@.  IC\,443 is in a complex region of
interstellar space, and the diffuse emission is thought to the product
of multiple supernova events (AA94).  The proper
motion problem is circumvented if the hard X-ray feature represents the
site of the most recent explosion which we can take from Chinese
records to have occurred in A.D. 837 (WAHK)\@.  If that were the case,
then it is surprising that this explosion has apparently not affected
the temperature or surface brightness distribution of the diffuse
emission.  The former suggests a more centrally located explosion; the
latter an explosion in the northeastern quadrant of the SNR.

On the other hand, the presence around the hard feature of many
interesting structures associated with collisions between the shock
front and concentrations of material suggests the hard X-ray emission
is related to them.  In particular, shock/cloud collisions can locally
enhance particle acceleration (JK93)\@.  We consider first whether
shock acceleration is a plausible source of the hard X-rays, and then
how the morphology of the feature might arise.

Dickel \& Milne (1976) found Faraday rotation measures in IC\,443 
of about 200 rad m$^{-2}$ where the field is along the line of sight, 
and close to zero
where it is perpendicular.  Taking the density and path length
measurements from the X-ray (PSSW),
we estimate the magnetic field strength $B$ to be:
\begin{equation}  \label{B_field.eq}
B \approx 1.23 \times
                  \left( \frac{ \rm 200 \, rad \, m^{-2} }
         {\rm 5 \, cm^{-3} \times 0.1\, pc} \right) \rm  \approx 500 \, \mu G 
\end{equation}
The synchrotron photon/electron energy relation therefore  can be
written:  \begin{equation} \label{photon_electron.eq}  E_{\rm photon}
\approx 5~{\rm keV} \times \frac{B}{\rm 500~ \mu G} \times
(\frac{E_{e}}{\rm 20~ TeV})^2 ~ , \end{equation} where $E_{\rm photon}$
is the observed photon energy produced as synchrotron radiation from an
electron at energy $E_{e}$\@.  So while radio emission ($E_{\rm photon}
\approx 10^{-9}$ keV) is produced by GeV electrons, production of 5 keV
X-rays requires the presence of $\sim$20 TeV electrons.

In attempting to explain the non-thermal component dominating the
X-ray emission from SN\,1006 as synchrotron emission from highly
relativistic electrons, Reynolds (1996) has shown that it is possible
to accelerate electrons (and ions) in SNR shocks to energies exceeding
100 TeV.  While his model is not strictly applicable here, some
aspects of it can be used to establish the plausibility of
shock accelerated electrons as the source of the hard X-ray emission
in IC\,443.

In the simplest models, the synchrotron spectrum is expected to be a
broken power law.  The various break frequencies correspond to electron
energies where two canonical times scales equate, such as the
synchrotron loss time and the acceleration time scale.  At each break,
the spectral index increases by about 0.5.  The electron diffusion time
is also important in determining the steepening of the spectrum, but
its influence is less straightforward to estimate.

In the IC\,443 hard feature, the difference between the radio and the
X-ray spectral index is approximately 1.1, suggesting the need for at
least two spectral breaks between the bands.  The time it takes for an
electron of energy E to lose half its energy via synchrotron radiation
in a magnetic field of strength B is given by:
\begin{equation}  \label{sync_loss.eq}
 \tau_{\rm loss} = \left( \frac{500~ \rm \mu G}{B} \right) ^2 
                      \left( \frac{20~ \rm TeV }{E_e} \right)
                       \times 2.5 \rm ~years \: ,
\end{equation}
which is approximately the light crossing time of the HXF\@.  Using
Equations \ref{sync_loss.eq} and \ref{photon_electron.eq}, and assuming
a SNR age between 1000 and 5000 yr (WAHK, AA94),
we find that the break frequency where the synchrotron loss time
becomes comparable to the age of the SNR occurs in the far infrared.
Additionally, by assuming a shock velocity on the order of 1000 km
s$^{-1}$, we find the break frequency associated with the equivalence
between the acceleration and loss times is in the hard ultraviolet
band.  The frequencies at which these two spectral breaks occur
supports the idea that the flat radio and the hard X-ray spectra are
both produced by shock acceleration.

While Reynolds' (1996) formalism facilitates the plausibility argument
above and is especially suggestive with regard to the ridge region, his
models for producing X-rays from shock accelerated electrons from a
large segment of the SNR shell clearly do not apply to the HXF in
IC\,443.  A different mechanism must be operating there.  Such a
mechanism has been suggested by JK93\@.

  Figure \ref{figure_discussion.fig} shows BGBW's H$_{2}$ image of
shocked molecular gas.  This shows that the molecular cloud has a
clumpy ring structure exterior to the site of the supernova event.
Dickman {\it et al.}\ (1992) found that if the ring is circular, it
must be inclined $\sim 51\arcdeg$ from the line of sight and have a
diameter of $\sim 9$ pc.  Doppler shift information shows the southern
edge of the ring (near our HXF) expanding toward us, and the northern
edge away -- implying that the ring is now expanding at a deprojected
velocity of 25 km/s.  While, we agree with this picture in principle,
we point out that to second order the ring is not circular; most
noticeably there is a section of negative curvature surrounding the HXF
(Fig.\ \ref{figure_discussion.fig})\@.  It is this change of curvature
that may create the HXF\@.

JK93 simulated cosmic-ray acceleration as shocks impact dense clouds --
which is the case here.  They found that after a shock/cloud impact,
particle acceleration is significantly enhanced in the ``tail shock''
emanating from the cloud in the direction of shock propagation.  Their
results suggest that each clump in the molecular ring could produce
enhanced shock acceleration in a long tail behind it.  Because of the
negative curvature of one section of the ring, the flow will be
refracted, thereby ``focusing'' the enhanced shock acceleration at the
hard X-ray feature where the tail shocks overlap.   This interaction
between these tail shocks will both increase the effective shock
velocities and provide a complicated structure which will require
further study.

The total hard flux from the diffuse emission along the shell (the
ridge) is about half that from the HXF, but the ridge may extend out of
the field of view.  It corresponds to an arc bright in H$_2$, and radio
continuum.  While we cannot yet be certain that the emission in
non-thermal, the higher ambient density as indicated by the infrared
line emission argues that the temperature of shock-heated material
should be lower than elsewhere in IC\,443 -- not higher.  The arguments
made above for the HXF against emission mechanisms other than
synchrotron hold for the ridge region as well.  While we withhold
judgment until a higher quality spectrum can be obtained, we suggest
that we are seeing evidence here as well for TeV electrons.  Whether
these arise from the JK93 mechanism or are more closely related to the
forward shock is still an open question.

\section{Conclusion \label{conclusion.sec}}

  In this paper we have presented evidence for a localized region of
particle acceleration to electron energies over 20 TeV, within the
shell type supernova remnant IC\,443\@.   Because of the unique
positions of the hard X-ray feature and shocked clouds, we believe that
the HXF is caused by the focusing of tail shocks around each molecular
cloud, thus enhancing particle acceleration in this location.    In
addition we have found a ridge of hard emission which in coincident
with both the radio-synchrotron shell and shocked molecular gas, but
whose spectral parameters are poorly constrained.

  More observational and theoretical research is needed to fully
understand this phenomenon.  Multi-frequency and polarization
observations with the VLA will allow us to both measure the feature's
spectral index and look for other indicators of on-going particle
acceleration.   A deep single-dish radio observation of this area is
necessary to verify that the HXF is indeed non-plerionic.  The most
important future observation will be to observe the full extent of the
ridge and the shocked molecular gas with ASCA to better characterize
the spectrum and image the full extent of the hard non-thermal
emission.

  On the theoretical side, this discovery provides a nice opportunity
  to model particle acceleration and cloud/shock interactions with
significant observational constraints.

\acknowledgements

We thank the following people for contributing to these results:
Richard Mushotzky for bringing the hard feature to our attention;
Michael Burton for kindly providing his H$_2$ emission line image in
digitized form; Steve Reynolds and Tom Jones for their insights into
shock acceleration; Matthew Baring and Ocker de Jager for their
thoughtful comments and suggestions.

\pagebreak[3]

\end{document}